%% file: proc_lat2000.tex
\documentstyle[twoside,fleqn,espcrc2,psfig]{article}
\input{newcommand.tex}
\title{Differential decay rate for $\Bpi$ semileptonic
  decays\thanks{Talk presented by T.~Onogi.}
  }
\author{
  JLQCD Collaboration: 
  S.~Aoki\address{
    Institute of Physics,
    University of Tsukuba, Ibaraki 305-8571, Japan}
  R. Burkhalter$^{\rm a, }$\address{
    Center for Computational Physics,
    University of Tsukuba, Tsukuba, Ibaraki 305-8577, Japan},  
  M. Fukugita\address{
    Institute for Cosmic Ray Research,
    University of Tokyo, Kashiwa 277-8582, Japan},
  S.~Hashimoto\address{
    High Energy Accelerator Research Organization (KEK),
    Tsukuba, Ibaraki 305-0801, Japan},  
  K-I.~Ishikawa$^{\rm d}$,  N.~Ishizuka$^{\rm a,b}$,
  Y.~Iwasaki$^{\rm a,b}$, K.~Kanaya$^{\rm a,b}$, T.~Kaneko$^{\rm d}$,
  Y.~Kuramashi$^{\rm d}$, M.~Okawa$^{\rm d}$, 
  T.~Onogi\address{
    Department of Physics, Hiroshima University,
    Higashi-Hiroshima, Hiroshima 739-8526, Japan}, 
  S.~Tominaga$^{\rm b}$, N.~Tsutsui$^{\rm d}$, A.~Ukawa$^{\rm a,b}$, 
  N.~Yamada$^{\rm d}$, 
  T.~Yoshi$\acute{e}$ $^{\rm a,b}$
  }
\begin{document}
\begin{abstract}
  We present our study on $B\rightarrow\pi l\nu$
  semileptonic decay form factors using NRQCD action for heavy quark.
  In the analysis, we use the form factors 
  $f_1(v\cdot k_{\pi})$ and $f_2(v\cdot k_{\pi})$ defined in the
  context of heavy quark effective theory by
  Burdman~\textit{et al.}.
  Since the NRQCD action for heavy quark respects the
  heavy quark symmetry, our results are described by the
  HQET form factors most naturally.
  From a quenched lattice QCD simulation at $\beta$=5.9 on
  a $16^3\times 48$ lattice, we obtain the form factors
  $f_1(v\cdot k_{\pi})$ and $f_2(v\cdot k_{\pi})$, and find
  that their $1/m_B$ correction is small.
  The limit of physical heavy and light quark masses can be
  reached without introducing any model function, and we
  obtain a prediction for the differential decay rate
  $d\Gamma/dq^2$.
  We also discuss the soft pion limit of the form factors.
\end{abstract}
\maketitle

\section{Introduction}
Model-independent calculation of the $B$ meson semileptonic
decay form factors is essential for extracting $|V_{ub}|$,
which is one of the most poorly determined elements of the
CKM matrix.
While the experimental results are being improved by the high
statistics data from the B factories, there are also
theoretical progresses 
in lattice QCD\cite{Maynard_lat00,Becirevic_lat00,Shigemitsu_lat00}, 
which has a potential
of making a precise prediction of the form factors near zero
recoil.

The success of lattice QCD for heavy-light physics
especially in the decay constant may be attributed to 
the use of the heavy quark (or non-relativistic) effective
theory (HQET).
The effective lattice actions constructed as a systematic
expansion in $1/m_Q$ provide a control over the systematic
errors from the large heavy quark mass. 
Furthermore, by identifying a quantity which has a
well-defined heavy quark mass limit using the HQET, one may
consider an expansion in $1/m_B$ and study the effect of
finite heavy quark mass in a systematic way.
Such a quantity for the decay constant is $f_B\sqrt{m_B}$,
for which many lattice calculations found that the $1/m_B$
correction is significant. 

In this study we take the same strategy for 
$B\rightarrow\pi l\nu$ semileptonic decay and use 
HQET motivated form factors $f_1(\vk)$ and $f_2(\vk)$
proposed by Burdman \textit{et al.} as
\cite{Burdman_et_al_94}
\begin{eqnarray}
\lefteqn{
  \frac{2}{\sqrt{m_B}} 
    \langle \pi(k_{\pi})|V^{\mu}|B(p_B) \rangle
    }
  \nonumber\\
  & = &
  f_1(\vk) v^{\mu} + 
  f_2(\vk) \frac{k_{\pi}^{\mu}}{v\cdot k_{\pi}},
  \label{eq:form_factor_def}
\end{eqnarray}
where $k_\pi$ and $p_B$ are four-momenta of $\pi$ and $B$
mesons in the external states, and $v^\mu = p_B^\mu/m_B$ is
a four-velocity of the $B$ meson.
The form factors $f_1(\vk)$ and $f_2(\vk)$ are well-defined
in the $m_B\rightarrow\infty$ limit, and it is natural to
consider a $1/m_B$ expansion around that limit.
We also note that $f_1(\vk)$ and $f_2(\vk)$ are implicit
functions of light quark mass $m_q$.
The relation to the conventional form factors $f^+(q^2)$ and
$f^0(q^2)$ is given by
\begin{equation}
  f^+(q^2) =
  \sqrt{m_B}
  \left[
    \frac{f_2(v\cdot k_\pi)}{v\cdot k_\pi} +
    \frac{f_1(v\cdot k_\pi)}{m_B}
  \right],
  \label{eq:f+}
\end{equation}
\begin{eqnarray}
  \lefteqn{f^0(q^2) =}
  \nonumber\\
  & &
  \frac{2}{\sqrt{m_B}}
  \frac{m_B^2}{m_B^2-m_{\pi}^2}
  \biggl[
    f_1(v\cdot k_\pi) + f_2(v\cdot k_\pi) 
  \nonumber\\
  & &
  \left.
    -\frac{v\cdot k_\pi}{m_B}
    \left(
      f_1(v\cdot k_\pi) +
      \hat{k}_{\pi}^2 f_2(v\cdot k_\pi)
    \right)
  \right],
  \label{eq:f0}
\end{eqnarray}
where we define 
$\hat{k}_{\pi}^{\mu} = k_{\pi}^{\mu}/(v\cdot k_{\pi})$
and
$q^2 = m_B^2 + m_{\pi}^2 - 2m_B (v\cdot k_{\pi})$.

\section{Lattice calculations}
We have performed quenched simulations at $\beta$=5.9 on a 
$16^3\times 48$ lattice, for which the lattice scale
determined from the string tension is
$1/a$ = 1.64~GeV.
We use the NRQCD action including $O(1/m_Q)$ terms for heavy
quark and the $O(a)$-improved light quark action with
$c_{\mathrm{SW}}$ calculated at one-loop 
$c_{\mathrm{SW}}$ = 1.58.
For the heavy quark mass we take $aM_0$ = 5.0, 3.0, 2.1 and
1.3, which cover the $b$ quark mass.
The light quark mass parameters, and some of the heavy quark
mass, are the same as in our previous study for $f_B$
\cite{JLQCD_fB_2000}. 
We accumulated 2150 independent gauge configurations to
reduce statistical error for the signals with finite spatial
momenta.
Even with this large number of statistics, the signal for
heaviest heavy or lightest light quark is not clean enough
to extract ground state.

The matching of the heavy-light vector current
$V^{\mu}=\bar{q}\gamma^{\mu}Q$ is done at
one-loop by Morningstar and Shigemitsu
\cite{Morningstar_Shigemitsu_99} 
and by Ishikawa \textit{et al.} \cite{Ishikawa_et_al}, in
which the lattice operators involved are
$\bar{q}\gamma_0 Q$, 
$\displaystyle -\frac{1}{2M_0} 
 \bar{q}\gamma_0\mathbf{\gamma\cdot\nabla}Q$,
and
$\displaystyle -\frac{1}{2M_0}
 \bar{q}\mathbf{\gamma\cdot\stackrel{\leftarrow}{\nabla}}Q$
for the temporal component, and
$\bar{q}\gamma_k Q$, 
$\displaystyle -\frac{1}{2M_0}
 \bar{q}\gamma_k\mathbf{\gamma\cdot\nabla}Q$,
$\displaystyle -\frac{1}{2M_0}
 \bar{q}\mathbf{\gamma\cdot\stackrel{\leftarrow}{\nabla}}\gamma_0\gamma_k Q$,
$\displaystyle -\frac{1}{2M_0}
 \bar{q}\gamma_0\nabla_k Q$, 
and
$\displaystyle \frac{1}{2M_0}
 \bar{q}\stackrel{\leftarrow}{\nabla}\gamma_0 Q$
for the spatial components.
In this work, we use the $V$-scheme coupling $\alpha_V(q^*)$
with $q^*=1/a$ for the coupling constant.

In the measurement of the matrix element we take several
combinations of initial and final meson momenta.
The maximum momentum we can measure is (1,0,0) for both
initial and final momenta in the unit of $2\pi/(16a)$.
As a result we can roughly cover a region 0.5$\sim$0.7 for
$av\cdot k_{\pi}$, where the lower limit is given by a pion
mass $am_{\pi}$.
By taking an inner product with $v^{\mu}$ of both sides of
(\ref{eq:form_factor_def}), we extract a linear combination
of form factors
$f_1(v\cdot k_{\pi})+f_2(v\cdot k_{\pi})$.
Another convenient basis is $f_2(v\cdot k_{\pi})$, which is
obtained by taking a inner product with an unit vector
perpendicular to $v^{\mu}$.

\section{HQET form factor results}

\begin{figure}[tb]
  \leavevmode\psfig{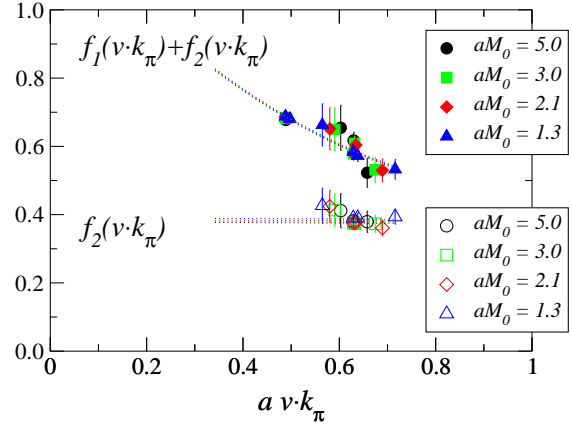}
  \caption{The HQET motivated form factors in unit of
    $a^{1/2}$ for four different heavy quark mass.
    }
  \label{fig:f1f2.k1}
\end{figure}

\begin{figure}[tb]
  \leavevmode\psfig{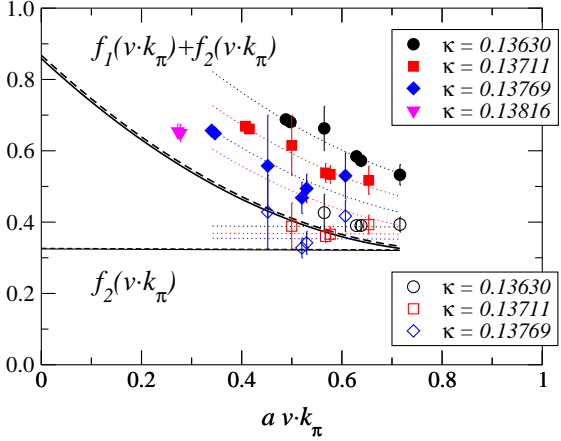}
  \caption{The HQET motivated form factors in unit of
    $a^{1/2}$ for four different light quark mass.
    }
  \label{fig:f1f2.mQ5}
\end{figure}

For the HQET motivated definition of the form factors
(\ref{eq:form_factor_def}), it is natural to expand the
heavy quark mass dependence in terms of $1/m_B$.
The light quark mass dependence can be expressed by a Taylor
expansion in $m_q$ when $v\cdot k_{\pi}$ is fixed.
Thus, we take the following form to fit the matrix
elements
\begin{eqnarray}
  \lefteqn{f_1(v\cdot k_{\pi}) + f_2(v\cdot k_{\pi})}
  \nonumber\\
  & = &
  A_0 + m_q A_1+ \frac{1}{m_B} A_2,
  \label{eq:fit_f1f2}
  \\
  f_2(v\cdot k_{\pi}) 
  & = &  
  B_0+ m_q B_1 + \frac{1}{m_B} B_2,
  \label{eq:fit_f2}
\end{eqnarray}
where the expansion coefficients $A_i$ and $B_i$ are
functions of $v\cdot k_{\pi}$. 
The expansions are truncated at first order, since we find
only mild mass dependences, as we shall see.
In practice, we parametrize $v\cdot k_{\pi}$ dependence of
the parameters by a Taylor expansion around $\vk=E_0$ to a
certain order.
We take $aE_0$ to be 0.5, and expand $A_0$ through quadratic
term, and $A_1$, $B_0$ through linear term.
Others are taken to be constant.

In Figure~\ref{fig:f1f2.k1} the form factors 
$f_1(v\cdot k_{\pi}) + f_2(v\cdot k_{\pi})$ and
$f_2(v\cdot k_{\pi})$ are plotted for four different values
of $aM_0$.
We find that $f_1(v\cdot k_{\pi}) + f_2(v\cdot k_{\pi})$
(filled symbols) has a significant negative slope in 
$v\cdot k_{\pi}$, while $f_2(v\cdot k_{\pi})$ (open symbols)
is consistent with constant.
Dotted curves in the plot represent the fit
(\ref{eq:fit_f1f2}) and (\ref{eq:fit_f2}).
It is interesting that there is almost no heavy quark mass
dependence, and therefore the $1/m_B$ correction to these
form factors is very small in contrast to the decay
constant, for which a large slope in $1/m_B$ was found in
lattice calculations.

At a fixed heavy quark mass ($aM_0$ = 1.3), the form factors
are plotted for different values of light quark mass in
Figure~\ref{fig:f1f2.mQ5}, where the fit curves are also
shown by dotted curves.
We observe a mild dependence on the light quark mass.
For $f_1(v\cdot k_{\pi}) + f_2(v\cdot k_{\pi})$, a large
fraction of the dependence comes from a trivial shift in
$v\cdot k_{\pi}$, which is
$E_{\pi}=\sqrt{m_{\pi}^2+\mathbf{k}_{\pi}^2}$ in the $B$
meson rest frame.
An extrapolation to the physical or chiral light quark limit 
using the fit formula (\ref{eq:fit_f1f2}) and
(\ref{eq:fit_f2}) is shown by dashed and solid curves
respectively.

\section{Differential decay rate}

\begin{figure}[tb]
  \leavevmode\psfig{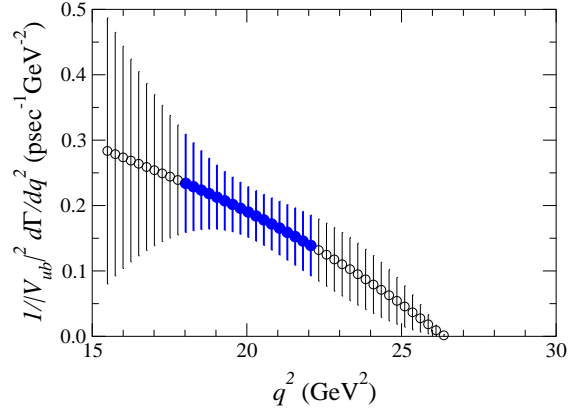}
  \caption{
    Differential decay rate for $B\rightarrow \pi l\nu$.
    }
  \label{fig:differential_decay_rate}
\end{figure}

From the fit results for the HQET form factors we may obtain
the usual form factors $f^+(q^2)$ and $f^0(q^2)$ using
(\ref{eq:f+}) and (\ref{eq:f0}).
The $q^2$ region where our lattice data is available is
18--22~GeV$^2$. 
The differential decay rate $d\Gamma/dq^2$ is then obtained
through 
\begin{equation}
  \label{eq:differential_decay_rate}
  \frac{d\Gamma}{dq^2} =
  \frac{G_F^2 |\mathbf{k}_{\pi}|^3}{24\pi^3}
  |V_{ub}|^2 |f^+(q^2)|^2,
\end{equation}
which is plotted in Figure~\ref{fig:differential_decay_rate}.
We note that only the results in the region $q^2$ =
18--22~GeV$^2$, which is given by filled symbols, are
obtained by interpolating the simulation data in 
$v\cdot k_{\pi}$, while the results outside the region
involve an extrapolation of $A_i$ and $B_i$ in 
$v\cdot k_{\pi}$.
Therefore, we should always keep in mind that the latter are
subject to systematic errors from the choice of fitting
functions. 

The error bars in the plot show the statistical error only. 
Systematic errors, such as the discretization effect 
$O((a\mathbf{k}_{\pi})^2)$ or the perturbative error
$O(\alpha_s^2)$, are under investigation.

\section{Form factors in the chiral limit}

\begin{figure}[tb]
  \leavevmode\psfig{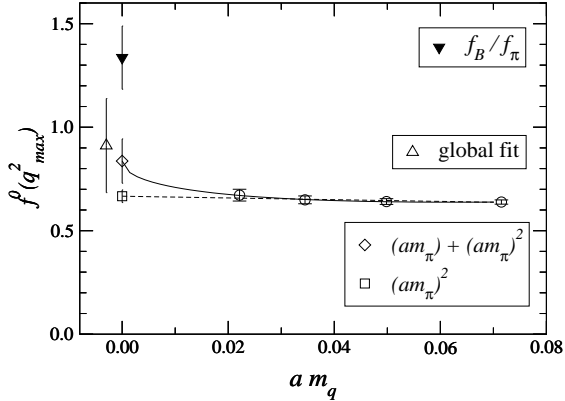}
  \caption{
    Soft pion limit of 
    $f_0(q^2_{max})$ =
    $2/\sqrt{m_B} [f_1(v\cdot k_{\pi}) + f_2(v\cdot k_{\pi})]$. 
    The dashed line is a linear fit in $(am_{\pi})^2$, while
    the solid curve includes the term $(am_{\pi})$.
    A result of the fit (\ref{eq:fit_f1f2}) is given by an
    open triangle, which should be equal to $f_B/f_{\pi}$
    (filled triangle) in the soft pion theorem.
    }
  \label{fig:soft_pion_limit}
\end{figure}

Despite the uncertainties in the extrapolation 
$v\cdot k_{\pi}\rightarrow 0$, it is important to study the
form factor in the soft pion limit 
($m_{\pi}$, $k_{\pi}\rightarrow 0$), 
since the HQET and chiral symmetry predict a relation
\begin{equation}
  f_1(0)+f_2(0) 
  = \frac{\sqrt{m_B}}{2} f^0(q^2_{max}) 
  = \frac{\sqrt{m_B}}{2} \frac{f_B}{f_{\pi}},
\end{equation}
where $f_B$ is the heavy-light decay constant.
In Figure~\ref{fig:soft_pion_limit} we compare the result of
the fit (\ref{eq:fit_f1f2}) shown by an open triangle with
the lattice calculation of $f_B/f_{\pi}$ (filled triangle)
\cite{JLQCD_fB_2000}.
Since the systematic errors, such as the uncertainty from
$v\cdot k_{\pi}$ extrapolation and higher order perturbative
corrections, are not yet included in the error bar, we
consider that they are in reasonable agreement. 
We also plot two extrapolations of 
$f_0(q^2_{max})$ =
$2/\sqrt{m_B} \left[f_1(v\cdot k_{\pi}) + f_2(v\cdot k_{\pi})\right]$
with the usual linear form in $(am_\pi)^2$ (dashed line) and
with a quadratic fit in $(am_\pi)$ (solid curve).
Since the value of $v\cdot k_{\pi}$ varies in the
extrapolation, linear dependence in $m_{\pi}$ appears
implicitly and the usual linear extrapolation in $m_{\pi}^2$
(or $m_q$) is no longer justified \cite{Maynard_lat00}.
Although the effect of the linear term in $am_{\pi}$ is very
small and is only seen at the lightest quark mass, it raises
the soft pion limit for the quadratic fit and makes it
consistent with the result from extrapolation using the fit
(\ref{eq:fit_f1f2}). 

Near the zero recoil limit, the HQET predicts the $B^{\ast}$
pole dominance \cite{Burdman_et_al_94}
\begin{eqnarray}
  \lefteqn{f_2(v\cdot k_{\pi})} 
  \nonumber\\
  & &
  \rightarrow
  g_{BB^*\pi}\frac{f_{B^*}\sqrt{m_{B^*}}}{2f_{\pi}}
  \frac{v\cdot k_{\pi}}{v\cdot k_{\pi} + m_{B^*}-m_B},
\end{eqnarray}
where $g_{BB^*\pi}$ denotes the $B^*B\pi$ coupling in the
heavy-light meson chiral effective theory.
Since the hyperfine splitting $m_{B^*}-m_B$ is small
compared to $v\cdot k_{\pi}$, we can approximate its
functional form by a constant in our data region.
Then, we obtain 
$g_{BB^*\pi} (f_{B^*}\sqrt{a m_{B^*}}/2f_{\pi})$ = 0.32(18),
which gives $g_{BB^*\pi}$ = 0.27(15).
It agrees with the phenomenological value extracted from
$D^*\rightarrow D\pi$ decay 0.27(6) \cite{Stewart_98}. 

\section*{Acknowledgment}
This work is supported by the Supercomputer Project No.54
(FY2000) of High Energy Accelerator Research Organization
(KEK), and also in part by the Grant-in-Aid of the Ministry
of Education (Nos. 10640246, 10640248, 11640250, 11640294,
11740162, 12014202, 12640253, 12640279, and 12740133).
K.-I.I., T.K. and N.Y. are supported by the JSPS Research
Fellowship.

\end{document}

%% file: newcommand.tex
\newcommand{\Bpi}{B\rightarrow\pi l \nu}
\newcommand{\be}{\begin{eqnarray}}
\newcommand{\ee}{\end{eqnarray}}
\newcommand{\vk}{v\cdot k_{\pi}}



%% file: proc_lat2000.bbl
\begin{thebibliography}{99}
\bibitem{Maynard_lat00}
  C.M.~Maynard, in these proceedings.
\bibitem{Becirevic_lat00}
  D.~Becirevic, in these proceedings.
\bibitem{Shigemitsu_lat00}
  J.~Shigemitsu, in these proceedings.
\bibitem{Burdman_et_al_94}
  G.~Burdman, Z.~Ligeti, M.~Neubert and Y.~Nir,
  Phys. Rev. \textbf{D49} (1994) 2331.
\bibitem{JLQCD_fB_2000}
  K.-I.~Ishikawa \textit{et al.} (JLQCD collaboration), 
  Phys. Rev. \textbf{D61} (2000) 074501 .
\bibitem{Morningstar_Shigemitsu_99}
  C.~Morningstar and J.~Shigemitsu,
  Phys. Rev. \textbf{D59} (1999) 094504.
\bibitem{Ishikawa_et_al}
  K.-I.~Ishikawa \textit{et al.}, in preparation.
\bibitem{Stewart_98}
  I.W.~Stewart, 
  Nucl. Phys. \textbf{B529} (1998) 62.
\end{thebibliography}
